\documentclass[pre,twocolumn,tightenlines,preprintnumbers,amsmath,amssymb]{revtex4}

\usepackage{graphicx}

\usepackage{epsfig}

\begin{document}
\preprint{ Recent Res. Devel. Stat. Phys., 2 (2002) 83-106 }

\title{Multiparticle random walks}
\author{L. Acedo}
\author{S. B. Yuste}
\affiliation{%
Departamento de F\'{\i}sica, Universidad  de  Extremadura,
E-06071 Badajoz, Spain
}%


\maketitle

\newpage

\section*{ABSTRACT}

An overview is presented of recent work on some statistical problems on multiparticle random walks. We consider a Euclidean, deterministic
fractal or disordered lattice and $N \gg 1$ independent random walkers initially ($t=0$)
placed onto the same site of the substrate.
Three classes of problems are considered:
(i) the evaluation of the average number $\langle S_N(t) \rangle $ of
distinct sites visited ({\em territory explored}) up to time $t$ by the $N$ random walkers,
(ii) the statistical description of the first passage
time $ t_{j,N}$ to a given distance of the first $j$ random walkers ({\em order statistics of exit times}), and
(iii)  the statistical description of the time $\mathbf{t}_{j,N}$ elapsed until the first $j$ random walkers are trapped when a Euclidean lattice is randomly occupied by a concentration $c$ of traps ({\em  order statistics of the trapping problem}).
Although these problems are very different in nature,  their solutions share the same form of a series in  $\ln^{-n}(N) \ln^m \ln (N)$ (with $n\geq 1$ and $0\leq m\leq n$) for $N \gg 1$. These corrective terms contribute substantially to the statistical quantities even for relatively large values of $N$.

\section{INTRODUCTION}
\label{sect_1}
It is now a commonplace in the history of physics to cite the works of the botanist Robert Brown (circa 1827) on the stochastic  movement of pollen grains suspended in water as the starting point in our understanding of the microscopic basis of diffusion.
In the ``annus mirabilis'' of 1905, Einstein
gave a successful theoretical explanation of the phenomenon in terms of an atomistic
theory which was later brilliantly confirmed by the experiments of Perrin
and served to remove the reluctance toward the atomic hypothesis that scientists as eminent as
Ostwald and Mach still entertained \cite{Einstein}. It was in this context that the theory of a
single random walker emerged. In the simplest random walk model a particle occupies
a site of a lattice (regular, fractal or disordered) and performs a jump to a
randomly selected nearest neighbour of that site every time step \cite{HughesWeiss}. Random
jumps are an effective way of simulating the net fluctuating force experimented by the
Brownian particle suspended in the liquid.  This discrete definition of
the random walk has gained widespread acceptance since the availability of computers
because, as are all lattice and automata models, it is especially suitable for computer
simulation. Random walks are also a way of describing fluctuations in the diffusion
process that are completely smoothed out  when we take the continuous limit represented
by diffusion equations.
Random walks have also been a very useful tool
in the study of transport in such disordered media  as fractured and porous rocks, silica
aerogels and percolation clusters \cite{SHDBA}, substituting more phenomenological
approaches \cite{Scheidegger}, and have also been used as topological models of
polymers (the so-called self-avoiding random walks \cite{BundeHav}).

The single random walker statistical problems have been the subject of intense
research since the beginning of past century and constitute now, in many areas, an almost closed discipline extensively  treated in general references \cite{HughesWeiss}.
However,  the generalization of these problems to the case of $N > 1$ interacting \cite{Interacting} or independent \cite{Larral1,Havlin,Shlesinger,fewN,BD,PRLYus,YusLin,DK,PREeucl,JPAYusAc,PREfract,YusAcLin,KR,HalfLine,TrapOrder} random walkers have  only really started to be considered in detail  in the last decade.
It may seem strange to the layman in random walk theory that there are any reasons for such a
late study of the multiparticle case, especially in the absence of random walker
interactions, because in other fundamental physical models composed of independent
particles, such as the ideal gas, the quantities of interest are obtained as simple
averages over the mechanical properties of single particles.
However, there are some quantities defined in a multiparticle random walk that can not be analyzed in terms of the single walker theory even when the walkers are independent! Examples are the number of distinct sites visited by a set of $N$ independent random walkers, $S_N(t)$, or the arrival time of the $j$-th particle of a set of $N$ independent random walker at a given border, $t_{j,N}$, neither of which yield to that simple approach.
Fortunately, these problems are tractable in the limit of large $N$ by resorting
to asymptotic \cite{PRLYus,YusLin,DK,PREeucl,JPAYusAc,PREfract,YusAcLin,HalfLine,TrapOrder}
or Tauberian and Abelian techniques \cite{HughesWeiss,BundeHav,Larral1,Havlin}. The complexity of these problems is a consequence of the superposition of the trails of the random walkers in the case of the territory problem and the competition between random walkers in the case of the order statistics.
Hence  every random walker has an influence on  the result which is indirectly correlated with the rest of the random walkers'  influences in spite of the absence of direct interactions between them.

Interest in multiparticle diffusion problems has been rekindled lately by  the development
of experimental techniques allowing the observation of events caused by single
particles of an ensemble \cite{SingMol,weitzlab}. These are powerful techniques aimed at the study
of local conditions (mechanical response, viscoelasticity) inside such complex structures
as fibrous polymers, the intracellular medium, etc.,  which determine
the behaviour of molecular motors and the rate of biochemical reactions. These new
research tools for the analysis of soft disordered media also demand a
better understanding of the statistical problems associated with mesoscopic samples
of Brownian particles.
\section{STATEMENT OF THE PROBLEMS AND DEFINITIONS}
\label{sect_2}

\subsection{Substrates}
We will consider three different classes of substrates in the following sections: (i) $d$-dimensional Euclidean
lattices ($d=1,2,3$), (ii) deterministic fractals (in particular, the two-dimensional Sierpinski gasket) and (iii) stochastic fractals (in particular, the two-dimensional incipient percolation aggregate embedded in the square lattice).

Euclidean lattices are reminiscent of crystalline structures and are the most customarily studied media,  so that the solution of our problems in this case is fundamental.
Fractal lattices  are good models of disordered substrates and have
been used extensively in the analysis of diffusion and transport properties in
these media.
In contrast with classical diffusion, the mean-square displacement of a random walker  in fractal
lattices is given by $\langle r^2 \rangle \sim 2 D t^{2/d_w}$, where $d_w > 2$ is the anomalous diffusion exponent ($d_w=2$ for classical diffusion) and $D$ is the diffusion coefficient.
This slowing down of the transport has also been observed in disordered media, which consequently supports the use
of fractals as models of real disordered materials \cite{SHDBA}.
Deterministic fractals are constructed by iteration of an unvariying rule \cite{SHDBA,BundeHav},
starting with a seed and successively applying the same iterator.
These kinds of fractal are only
mathematical idealizations but they have the obvious advantage of being suitable
for the use of exact renormalization techniques in the calculation of many statistical
quantities \cite{PRLYus,Broeck}.
Unfortunately, stochastic fractals do not allow the application of these
techniques because they are the result of some random process.
However, they are also closer to real disordered media which share their statistical-fractal structure \cite{SHDBA,BundeHav,Stauffer,Fractals}.

\begin{figure}
\includegraphics{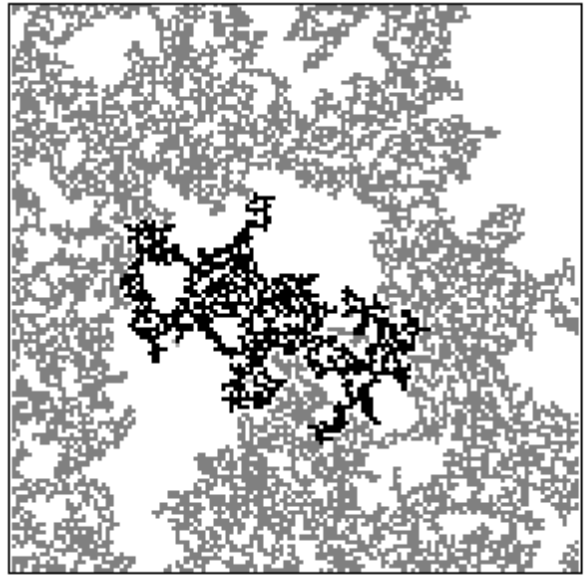}
\caption{\label{figagreg}
Territory explored (black points) by $N=1000$ random walkers diffusing on an two-dimensional percolation aggregate (gray points) after one thousand steps ($t=1000$). The starting point is at the center of the figure.}
\end{figure}

The most widely known stochastic fractal used in theoretical and simulation studies is the
percolation aggregate \cite{SHDBA,Stauffer,Fractals}.
A percolation aggregate is a cluster of sites in a regular lattice that are assumed to be connected by bonds
between nearest-neighbours.
This cluster is constructed by the random filling  with probability $p$ of the nodes of a regular lattice.
At a certain critical concentration $p_c$ of occupied sites an infinite cluster, called the incipient percolation aggregate, appears. The incipient percolation aggregate is a fractal with fractal dimension $d_f=91/48$ if embedded in two dimensions and $d_f \simeq 2.5$ if constructed on a three-dimensional lattice.
In our simulation we considered only the two-dimensional incipient percolation aggregate on the square lattice
constructed by the standard Leath method \cite{Fractals,Leath} using the value
$p=p_c=0.5927460\cdots$ corresponding to site percolation in the square lattice \cite{Stauffer}.
In Fig. \ref{figagreg} we show a portion of an incipient percolation aggregate in a two-dimensional square lattice with $L \times L$ sites.
Multiparticle random walk problems with independent random walkers have been posed and studied on all of these media.
The ones to be considered in this review are described below.

\subsection{The number of distinct sites visited. Territory explored}
A set of $N$ random walkers are placed on the same site of a lattice at $t=0$.
As time runs the random walkers move independently, jumping
randomly from the site they occupy to any of its nearest-neighbours and imprinting every
site they visit. We are concerned with those sites visited by any of
the random walkers. Successive visits are not relevant to register a site as
visited. The number $S_N(t)$ of the imprinted sites, that is, the number of distinct sites
visited at time $t$ is the territory explored by the diffusing random walkers.
This is the magnitude whose statistical distribution we are interested in.

The case $N=1$ was posed at the beginning of the 1950s by Dvoretzky and Erd\"os \cite{DE} and has been thoroughly studied since then
\cite{HughesWeiss}. The problem was taken up again by Larralde {\em et al.}
\cite{Larral1,Havlin} who systematically treated the multiparticle ($N > 1$) version.
In these pioneering works the general features of the solution in the limit $N \gg 1$
were unveiled, and they found three time regimes as follows
\begin{equation}
\langle S_N(t) \rangle \sim \left\{ \begin{array}{ll}
\text{const}\times t^d, &  t \ll t_{\times} \\
\noalign{\smallskip}
\text{const}\times  t^{d/2} \ln^{d/2}\left( x \right), & t_{\times} \ll t \ll
t_{\times}^{'} \\
\noalign{\smallskip}
N \langle S_1(t)  \rangle , &  t_{\times}^{'} \ll t
\end{array} \right.
\end{equation}
where $x=N$ for $d=1$, $x=N/\ln t$ for $d=2$ and $x=N/\sqrt{t}$
for $d=3$ \cite{Larral1,Havlin}. The behaviour of the territory
covered by a single random walker, $\langle S_1(t) \rangle $, is also well known:
$\langle S_1(t) \rangle  \sim \text{const}\times  t^{1/2}$ for $d=1$, $\langle S_1(t) \rangle  \sim \text{const}\times  t/\ln t$ for $d=2$, and $\langle S_1(t) \rangle  \sim \text{const}\times  t$ for
$d=3$. The existence of these three regimes is easily explained:

\paragraph{Regime I.} In this case there are so many particles at every site that all the nearest neigbours of the
already visited sites are reached at the next step. It is clear that the territory
covered by the random walkers grows as the volume of a hypersphere of radius $t$, $\langle S_N\rangle \sim \text{const}\times t^d$. The
crossover time from regime I to regime II, $t_{\times}$, is simply derived if we take
into account that regime I must break when the number of particles on the outer
zone of the territory visited is of order $1$.
For very short times the number of
particles on the outer visited sites decreases
as $N/z^t$, where $z$ is the coordination number of the lattice, and, thus, the overlapping regime will break approximately when $N/z^t \simeq 1$ or,  equivalently, when $t_{\times} =\mathcal{O}(\ln N)$.

\begin{figure}[t]
\includegraphics[width=8cm]{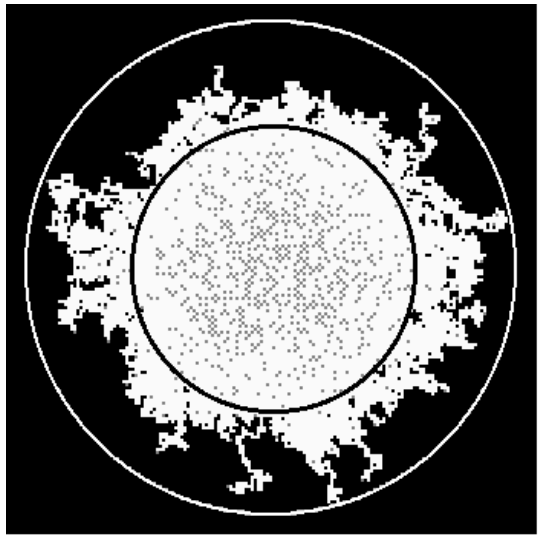}
\caption{\label{Territory}
A snapshot of the set of sites visited by $N=1000$ random walkers
on the two-dimensional lattice.
The visited sites are in white, the unvisited ones
are in black and the internal gray points are the random walkers. The outer
white circle is centered on the starting point of the random walkers
and its radius is the maximum distance from that point reached by any walker
at the time the snapshot was taken. The internal black circle is
concentric with the former but its radius is the distance between the origin
and the nearest unvisited site.}
\end{figure}

\paragraph{Regime II.} This is the most interesting regime. The simple regime I has broken and the random walkers move diffusively so that the radius of the territory explored grows as $t^{1/2}$. Then the territory explored $\langle S_N(t) \rangle $ is given, essentially, by a volume that grows as  $t^{d/2}$ modified by a factor
that depends on $N$. The explored region is divided into an inner
hyperspherical core and a corona of dendritic nature
characterized by filaments created by the random walkers
wandering in the outer regions (Fig. \ref{Territory}). The overlapping of the trails of the random
walkers make the problem non-trivial \cite{Larral1,Shlesinger}.
In Refs.\  \cite{PREeucl,PREfract}, asymptotic techniques were used to obtain the prefactor and corrective terms of the main term of $\langle S_N(t)\rangle $.

\paragraph{Regime III.}
Particles are very far from each other and their trails (almost)
never overlap so that $\langle S_N\rangle \sim N \langle S_1 \rangle $. The territory explored by the set of $N$ random
walkers is the sum of the territories explored by $N$ single random walkers. The
crossover time from regime II to regime III is $t_{\times}^{'} =\mathcal{O}(e^N)$ for $d=2$ and
$t_{\times}^{'} =\mathcal{O} (N^2)$ for $d=3$ \cite{Larral1,PREeucl}.
This regime never appears if the spectral dimension $d_s=2 d_f/d_w$ of the substrate
is $d_s < 2$. In particular, regime III is never reached in one-dimensional
lattices.

The territory problem was also studied for {\em fractal} lattices with
$d_s < 2$ by Larralde {\em et al.} \cite{Larral1} and Havlin {\em
et al.} \cite{Havlin} who proposed the expression $\langle S_N \rangle \sim
\text{const}\times t^{d_s/2} \left(\ln N\right)^{d_f/u}$ for the regime II with $N
\gg 1$, where $u=d_w/(d_w-1)$.
More recently, Dr\"ager and Klafter \cite{DK} have also analyzed this problem using scaling
arguments finding that $\langle S_N \rangle =\mathcal{O} \left[ t^{d_s/2} (\ln
N)^{d_\ell/v}\right]$ for $t_{\times} \ll t$, where
$v=d_w^\ell/(d_w^\ell-1)$ and  $d_w^\ell=d_w/d_{\text{min}}$ is
the chemical-diffusion exponent \cite{SHDBA,BundeHav,Stauffer}.
Of course, the two predictions agree for those media, such as
Sierpinski gaskets, for which $d_{\text{min}}=1$ but disagree for those with $d_{\text{min}}\neq 1$, such as the
two-dimensional and three-dimensional incipient percolation aggregates  for which $d_{\text{min}} \simeq 1.15$  and
$d_{\text{min}} \simeq 1.33$, respectively.
It is notable that both Havlin {\em et al.} and Dr\"ager and Klafter supported their
conjectures by comparison with simulation results obtained for two- and three-dimensional percolation aggregates.
We have shown that their collapsing plots are not conclusive because in these plots the influence of the  large logarithmic corrective terms is not considered \cite{PREfract}.

\subsection{ Order statistics of exit times}
Order statistics is a relatively young field of mathematical statistics \cite{ABN}. Its objective is
the ordering of sequences of a set of random variables.
Ours is also a problem of order statistics for the sequence of first-passage times (exit times) of a set of $N$ independent random walkers, all starting from the same site at the same time, to  a ``spherical'' boundary or radius $z$ in Euclidean and fractal media.
In particular, we are interested in the calculation of the moments $\langle t_{j,N}(z) \rangle$ of the
$j$th passage time $t_{j,N}(z)$. This is the time taken by the $j$th random walker of a set of $N$
to first reach a given distance $z$. Some early results concerning the order statistics
of a set of random walkers on Euclidean lattices were obtained by Lindenberg {\em et al.} \cite{Lindenberg}
and Weiss {\em et al.} \cite{fewN}. Asymptotic results ($N \gg 1$) for finitely
ramified fractals were obtained much later by one of us \cite{PRLYus} using
renormalization techniques developed by van den Broeck \cite{Broeck}. This kind
of fractal  has the property that,  by cutting a finite number of bonds, a portion of the lattice with certain number of generations becomes isolated from the rest of the lattice.
Curiously, rigorous asymptotic expansions for Euclidean lattices of arbitrary
dimension were only given later by Yuste {\em et al} \cite{YusAcLin}.

\subsection{Order statistics of the trapping problem.}
The ``trapping'' problem has for decades been one of  most widely studied the areas of
random walk theory \cite{HughesWeiss,SHDBA}. In its simpler version
we have a Euclidean lattice where a population of traps occupies the sites with
probability $c$. A single random walker starts moving from an empty site until it is
absorbed by one of the traps. The statistical quantity of interest is the survival
probability $\Phi_1(t)$ that the random walker is not trapped by time $t$. This
problem has its origin in Smoluchowski's theory of coagulation of colloidal
particles \cite{HughesWeiss,SHDBA,Hollander} and has been applied to many systems
in physics and chemistry such as trapping of mobile defects in crystals with
point sinks \cite{Beeler,Rosens,Damask}, the kinetics of luminescent organic
materials \cite{Rosens}, anchoring of polymers by chemically active sites
\cite{Oshanin} and atomic diffusion in glasslike materials \cite{Miyagawa}, among
others. The generalization of the trapping problem to $N$ independent random walkers
was only considered very recently by Krapivsky and Redner \cite{KR} who studied
a predator-prey problem in which a static prey or ``lamb'' is captured by one of
a set of $N$ diffusing predators or ``pride or lions'' in one dimension.  In their
problem the $N$ predators are placed initially at a given distance from the prey. Later, the
case was studied of a stochastic distribution of prey on the half-line \cite{HalfLine}.
Also very recently \cite{TrapOrder} an approach has been made to the evaluation of the $m$th moment $\langle t_{j,N}^m \rangle$ of the time $t_{j,N}$ elapsed until the first $j$ random walkers of a set of $N$ are trapped
in a $d$-dimensional Euclidean lattice populated by traps at a concentration $c$.
In Sec.\ \ref{trapping} we will discuss the solution to this problem.

\section{SURVIVAL PROBABILITY: ABSORBING TRAPS AND BOUNDARIES}
\label{survival}
The calculation of the average territory covered by $N\gg 1$ random walkers and the order statistics of exit and trapping times requires the previous knowledge of the {\em survival} functions, or their complementaries, the  {\em mortality}  functions for short times.
Suppose we have a circular boundary of radius $r$ in a lattice and a single
random walker starting at the center of this circle at $t=0$.
The probability that this diffusing particle has reached the distance
$z$ during the time interval $(0,t)$ is called the mortality function.
We will denote it as $h_B(z,t)$ where the subscript $B$
indicates that this quantity refers to the boundary and
$z$ denotes any distance defined on the substrate.
In stochastic fractal lattices it is common to define the chemical distance $\ell$
as the shortest path measured along the lattice bonds \cite{SHDBA}.
Between this chemical distance and
the Euclidean distance there exists a scaling relation, $\ell
\sim \text{const}\times r^{d_{\text{min}}}$, with $d_{\text{min}}=1$ for Euclidean
and deterministic fractals (Sierpinski gasket, Given-Mandelbrot
curve, etc\ldots) and $d_{\text{min}} > 1$ for stochastic fractals
(for example, $d_{\text{min}}=1.14(2)$ for the two-dimensional incipient
percolation aggregate). Thus, in the case of stochastic fractal
lattices there are at least two independent ways of defining a
``spherical'' boundary as the set of sites with either constant $\ell$
(chemical boundary) or constant $r$ (Euclidean boundary) from the
origin. The anomalous diffusion coefficient $d_w^z=d_w/d_{\text{min}}$ appears in
Einstein´s relation, $\left\langle z^2 \right\rangle \sim 2 D
t^{1/d_w^z}$, for the average square distance traveled by the
random walker by time $t$, so that $d_w^z=d_w$ for $z=r$ and
$d_w^z=d_w^{\ell}$ for $z=\ell$.
The complementary of $h_B(r,t)$ is the survival
probability in the case of an absorbing boundary: $\Gamma_B(r,t)=1-h_B(r,t)$.
Similarly, we will define the mortality function of a single random walker
starting from the origin site of a lattice with a trap located on
site ${\bf r}$ as the probability that site ${\bf r}$ has
been visited by that random walker in the time interval $(0,t)$.
We will denote it as $h_P({\bf r},t)$ in order to distinguish it
from $h_B(r,t)$.

These functions have been studied through the
application of various techniques: renormalization in the case of
finitely ramified fractals \cite{PRLYus,Broeck,Martinez}, the
solution of the diffusion equation for Euclidean media
\cite{YusAcLin,Bidaux} and computer simulations
\cite{PREfract,fptagreg}.
 The resulting expression in the short-time (large-distance) limit  is
given by
\begin{equation}
\label{hbzt}
h_B(z,t) \sim A \xi^{-\mu v} e^{-c \xi^v} \left\{1+h_1 \xi^{-v}+\ldots\right\}\; ,
\end{equation}
with $\xi=z/\langle z^2\rangle^{1/2}=z/(\sqrt{2D}t^{1/d_w^z})\gg 1$ and $v=d_w^z/(d_w^z-1)$.
 The parameters $d_w^z$, $A$, $\mu$, $c$ and $h_1$ are listed in Table \ref{table1} for several lattices.

\begin{table*}
\caption{\label{table1}
Parameters appearing in the asymptotic expression of the
mortality function $h_B(z,t)$ of a random walker starting at the origin with
a {\em boundary} of trapping sites at a given distance, Eq.\
(\protect{\ref{hbzt}}) for four substrates: the symbol dSC refers
to the $d$-dimensional simple cubic lattice, Sd to the $d$-dimensional
Sierpinsky lattice, GM to the Given--Mandelbrot curve ($d_w^z=d_w$ for these cases) and A2E
(A2C) to the two-dimensional incipient percolation aggregate with
Euclidean (chemical) boundary. These parameters are analytical
for Euclidean lattices \protect\cite{YusAcLin,Bidaux}, numerical for the
deterministic fractal lattices \protect\cite{PRLYus,YusteJPA} and the
output of a numerical fit to simulation results for the
percolation aggregate \protect\cite{fptagreg}. In this case a $2000$
aggregates average and a time interval $[0,1000]$ were used.}
\begin{ruledtabular}
\begin{tabular}{ccccccc}
Case & $d_w^z$ &2D& $A$ & $\mu$ & $c$ & $h_1$ \\
\hline
dSC & 2 & 1&$\displaystyle\frac{2 (d/2)^{d/2-1}}{\Gamma(d/2)} $ & $1-d/2$& $d/2$ & $\displaystyle\frac{d-3}{2d}$ \\
S2 & $\ln 5/\ln 2$&1.05  & 2.46 & 1/2 & 0.981 & -0.56  \\
S3 & $\ln 6/\ln2$ &--  & 3.36 & 1/2 & 1.31  & -0.46  \\
GM & $\ln 22/\ln 3$&-- & 2.5 &  1/2 & 1.10  & -0.6   \\
A2E & 2.8&1.4 & 1.6 & -1.8 & 2.1 & -- \\
A2C & 2.4&1.2 & 1.1 & -0.4 & 1.1 & --
\end{tabular}
\end{ruledtabular}
\end{table*}

\begin{figure}
\includegraphics{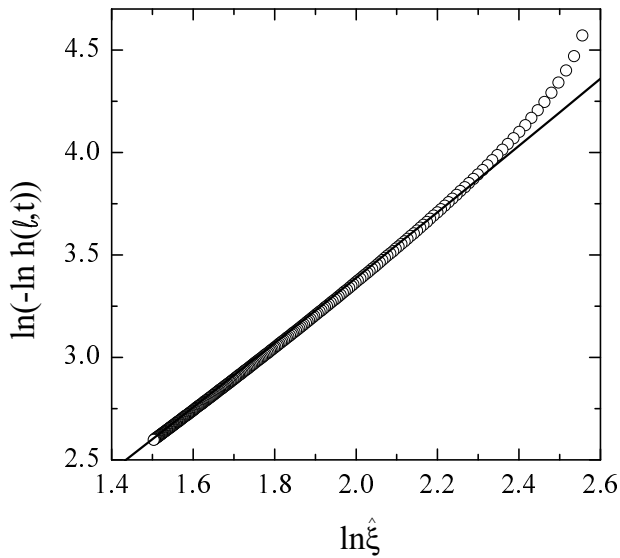}
\caption{\label{fighpt}
Plot of $\ln[-\ln h_P(\ell,t)]$ versus $\ln \xi$ averaged over $2000$
two-dimensional incipient percolation clusters. The trap was always placed at a
site a distance $\ell=80$ from the origin. The line represents the function
of Eq.\ (\protect\ref{hpzt}) with the parameters listed in Table \protect\ref{table2}.}
\end{figure}

It is known that the mortality function of a single random walker
starting at a distance $z$ from a lattice trap site is given by a
an equation formally identical to (\ref{hbzt}) in the case of Euclidean
lattices \cite{YusAcLin,Bidaux} for $\xi \gg 1$. This form
also encompasses the cases of deterministic and stochastic fractal lattices,
as simulation results have shown \cite{PREfract}. Hence, in general, we will
write
\begin{equation}
\label{hpzt}
h_P(z,t) \sim \hat{A} \xi^{-\hat{\mu} v} e^{-\hat{c} \xi^v} \left\{1+\hat{h}_1 \xi^{-v}+\ldots\right\}\; ,
\end{equation}
for  $\xi \gg 1$, and where $\hat{A}$, $\hat{\mu}$, $\hat{c}$ and $\hat{h}_1$ is a set of parameters characteristic of the lattice.
In Table \ref{table2} the values of these parameters are listed for several lattices.
In the absence of exact theoretical expressions for $h_P(z,t)$, these parameters are obtained from comparison with simulation results. Figure \ref{fighpt} shows an example. In this figure the theoretical short-time behaviour of $h_P(z,t)$ given by Eq.\ \eqref{hpzt} is compared to simulation results for the two-dimensional incipient percolation aggregate. One sees that the values of $d_w^z$, $A$, $\mu$, $c$ listed in Table \ref{table2} lead to good agreement between Eq.\ \eqref{hpzt} and the simulation results.

\begin{table*}
\caption{\label{table2}
The same as Table \protect{\ref{table1}} but for the
mortality function of a single random walker starting at a
distance $z$ from a trapping {\em site}, $h_P(z,t)$. Results for the
two-dimensional incipient percolation aggregate with a trap at a
given Euclidean distance from the origin are not shown because
the broadness of the distribution does not allow any reasonable
fit for the limited number of aggregates used in our simulations (See Fig.\
\protect\ref{histo}). The parameter $\tilde{p}$ is $[ 2 (2 D
\pi)^3/3 ]^{1/2} p({\bf 0},1)]$, where $p({\bf 0},1) \simeq
1.516386$ \protect\cite{HughesWeiss}. The chemical dimension
$d_\ell$ and the volume $V_0$ of a hypersphere of chemical radius $\ell=1$, are also
listed.}
\begin{ruledtabular}
\begin{tabular}{cccccccc}
Case & $d_w^z$ & $d_\ell$  & $V_0$ & $\hat{A}$ & $\hat{\mu}$ & $\hat{c}$ & $\hat{h}_1$ \\
\hline
1SC & 2 & 1 & 2 & $\sqrt{2/\pi}$ & 1/2 & 1/2 & -1 \\
2SC & 2 & 2 & $\pi$ & $1/\ln t$ & 1 & 1 & -1  \\
3SC & 2 & 3 & $3 \pi/2$ & $1/\tilde{p} \sqrt{t}$ & 1 & 3/2  & -1/3  \\
S2 & $\ln 5/\ln 2$ & $\ln 3/\ln 2$ &  3 & 0.61 &  1/2 & 0.98  & -0.56   \\
A2C & 2.4 & 1.65 &  1.1 & 1.0 & 0.8 & 1.05 & --
\end{tabular}
\end{ruledtabular}
\end{table*}

\begin{figure}
\includegraphics{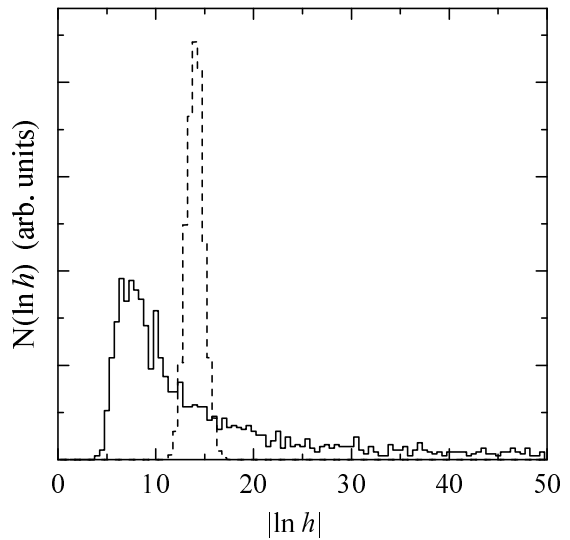}
\caption{\label{histo}
Plot of the histogram ${\cal N}(\ln h_P)$ versus $\vert \ln h_P \vert$
for the two-dimensional incipient percolation aggregate for fixed $r$ and $t$ (solid line)
and for fixed $\ell$ and $t$ (dashed line). The values are $r=30$, $\ell=80$, and
$t=1000$.}
\end{figure}

Finally, it is interesting to note that, for disordered media,  the territory explored and the order statistics of the exit times are better described in terms of the chemical distance
\cite{DK,PREfract,fptagreg} than in terms of the Euclidean distance.
The reason for the advantage of the chemical distance description is to be found in
the broadness of the distribution of the mortality function at fixed time $t$ and
Euclidean distance $r$ in comparison with the corresponding distribution for fixed chemical
distance $\ell$. Figure \ref{histo} shows the histogram for the values of $h_P(z,t)$ for the values $r=30$ and $\ell=80$ at $t=1000$ plotted from the results for
$2000$ percolation clusters. The same Euclidean distance in a disordered lattice
corresponds to very different chemical paths depending on the holes that may
block this path between the origin and the destination sites. Hence, the
mortality function should exhibit large deviations from one realization of
the fractal lattice to another.

\section{Territory explored}
\label{territory}
\subsection{Asymptotic expressions}
In this section we give the main and two first corrective terms of the asymptotic expression (a power series in $\ln^{-1} N$) of the territory explored $\langle S_N(t) \rangle $ by $N\gg 1$ random walkers, thus
going beyond the leading terms (without prefactor) discussed in Sec. \ref{sect_2}.
In the previous section we have defined the survival probability $\Gamma_P({\bf r},t)$ as the probability
that a site ${\bf r}$ has not been visited by a single random walker by time $t$. The average
number of distinct sites visited by $N$ independent random walkers, $\langle S_N(t) \rangle $, is then
simply related to $\Gamma_P({\bf r},t)$ as follows \cite{Larral1,Havlin}:
\begin{equation}
\label{SNt}
\langle S_N \rangle =\left\langle\sum \, \left\{ 1 - \left[\Gamma_P({\bf r},t)\right]^N \right\}\right\rangle \; ,
\end{equation}
where the sum runs over all the sites of the lattice.
One  must notice that there are two averages implicit in Eq. (\ref{SNt}):
(a) an average over all exploration experiments performed on the
same lattice, represented by the sum $\sum  \left\{ 1 - \left[\Gamma_P({\bf r},t)\right]^N \right\}$; and
(b) a second average
$\left\langle \ldots\right\rangle$ over all possible stochastic
lattices compatible with the generation rules.
Of course, in the case of
deterministic lattices (Euclidean, Sierpinski gaskets, etc\ldots)
only the first average is necessary.
As the histogram of the survival probability $\Gamma_P(z,t)$ is very narrow in stochastic fractal lattices when the trap is placed at fixed chemical distance  $z=\ell$ (see Fig.\ \ref{histo}), it is not difficult to see that Eq.\ \eqref{SNt} can be approximated by
\begin{equation}
\label{SNtl}
\langle S_N \rangle =\sum_{m=0}^\infty \left\{1-\left[\Gamma_P(\ell_m,t)\right]^N \right\}\left\langle n(m) \right\rangle\; ,
\end{equation}
where $\langle n(m) \rangle$ is the average number of sites separated from the origin
by a chemical distance in the range $[\ell_m=m \Delta \ell,\, \ell_m+\Delta \ell]$, $m=0,1,2,\cdots$
Evidently, Eq.\ (\ref{SNtl}) is exact when applied to Euclidean lattices and deterministic fractals and, for
the sake of generality, we will take it as the starting point of the subsequent derivations for all classes of substrates.

\begin{figure}
\includegraphics{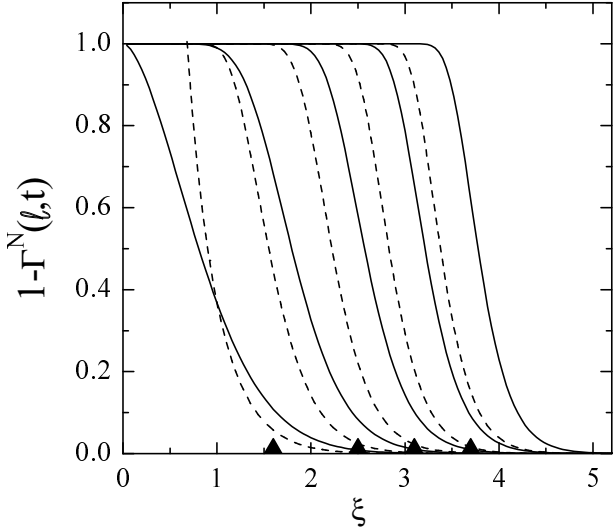}
\caption{\label{figHeav}
Function $1-[\Gamma_t(\ell)]^N$ versus $\xi=\ell/L(t)$  for (from left to right) $N=1, 10,100,1000$ and $ 10000$ where $\Gamma_t(\ell)=1-\xi^{-\mu v} \exp(-\xi^v)$,  $v=1.7$, $\mu=0$ (solid line) and $\mu=0.8$ (dashed line). We have not plotted the unphysical values that appear in the case with $\mu=0.8$ when $\xi$ goes to zero.
Notice the large influence of the subdominant power term $\xi^{-\mu v}$ on the value of $[\Gamma_t(\ell)]^N$ which  will be reflected in the value of $S_N(t)$.
The triangles mark the value of $(\ln N/c)^{1/v}\simeq \xi_{\times}$ for
$N=10$, $100$, $1000$, and $10000$.}
\end{figure}

First, we replace Eq. (\ref{SNtl}) by its continuum approximation
\begin{equation}
\label{SNtint} \langle S_N \rangle =\displaystyle\int_0^\infty\,\left\{ 1 -
\left[\Gamma_P(\ell,t)\right]^N \right\}
 d_\ell\, V_0\, \ell^{d_\ell - 1} d \ell \;  ,
\end{equation}
where $d_\ell$ is the chemical dimension of the fractal lattice
and $d V(\ell)=V_0\, d_\ell \, \ell^{d_\ell - 1} d \ell$ is the
average number of fractal sites placed at a chemical distance
between $\ell$ and $\ell+d\ell$ (values of $V_0$ for several lattices are given in table \ref{table2}). The asymptotic evaluation technique
for $\langle S_N \rangle $ is inspired in the behaviour of
$1-\left[\Gamma_P(\ell,t)\right]^N$ for a fixed time $t$. This
function is plotted in Fig. \ref{figHeav} for several values of
$N$ in the case of the two-dimensional incipient percolation
aggregate.
We observe that it approaches a unit step function
$\Theta(\ell-\ell_{\times})$ when $N \rightarrow \infty$,
$\ell_{\times}$ being a value that depends on $N$. For large $N$,
$[\Gamma_P(\ell,t)]^N$ is only non-negligible when
$\Gamma_P(\ell,t)$ is very close to $1$. This occurs in the limit
$\xi=\ell/L(t) \gg 1$ ($L(t)\sim \sqrt{2 D} t^{1/d_w^\ell}$ is the
root-mean square chemical distance traveled by a single random walker by time
$t$). On the other hand, $1-[\Gamma_P(\ell,t)]^N$ approaches rapidly the value
$1$ as $\xi$ decreases (large times and short-distance limit). Therefore, it is
clear that $1-[\Gamma_P(\ell,t)]^N \approx \Theta(\ell-\ell_{\times})$ is a reasonable
approximation which improves as $N$ increases. The threshold chemical distance
$\ell_{\times}$ can be reasonably defined as the distance at which $1-[\Gamma_P(\ell,t)]^N$
takes the value $1/2$, and from Eq.\ (\ref{hpzt}) we find that $1/2 \approx
N \hat{A} \xi_{\times}^{-\hat{\mu} v} \exp(-\hat{c} \xi_{\times}^v)$ and the following
approximation $\hat{c} \xi_{\times}^v \approx \ln N-\hat{\mu} v \ln \xi_{\times}+\ln 2 A$ for $\xi_{\times}$ ensues.
We then get $\hat{c} \xi_{\times}^v \approx \ln N-\hat{\mu} \ln \ln N+\ln A c^\mu+
\ln 2$ or
\begin{eqnarray}
\label{ltimes} \ell_{\times} &\approx& (2 D)^{1/2}
t^{1/d_w^\ell} \left(\frac{\ln N}{\hat{c}}\right)^{1/v} \nonumber \\
&& \times \left(1+\frac{1}{v} \frac{-\hat{\mu} \ln \ln N+\ln \hat{A}
\hat{c}^{\hat{\mu}}+\ln 2}{\ln N} \right).
\end{eqnarray}
The integration in Eq.\ (\ref{SNtint}), evaluated using the Heaviside step
function approximation for the quantity inside the bracets, trivially yields
$\langle S_N(t) \rangle  \approx V_0 \ell_{\times}^{d_\ell}$, and after inserting the expression
for $\ell_{\times}$ in Eq.\ (\ref{ltimes}) we finally arrive at the following
approximation for $\langle S_N \rangle $:
\begin{eqnarray}
\langle S_N \rangle  &\approx&  V_0 (2 D)^{d_\ell/2} t^{d_\ell/d_w^\ell}
\left(\frac{\ln N}{\hat{c}}\right)^{d_\ell/v} \nonumber\\
&&\times \left\{ 1+
\frac{d_\ell}{v}\; \frac{\ln 2 + \ln \hat{A}
\hat{c}^{\hat{\mu}}-\hat{\mu}\ln \ln N}{\ln N} \right\}.
\label{SNtapprox}
\end{eqnarray}
The elementary approach leading to Eq.\ (\ref{SNtapprox}) is not exact,  but the
dominant behaviour and the form of the first corrective term found (except for the
$\ln 2$ in the numerator) coincide with the prediction of a systematic and rigorous
improvement of the approach discussed above. This rigorous analysis
for $\langle S_N \rangle $ yields  \cite{PREeucl,PREfract}
\begin{equation}
\langle S_N \rangle \sim \widehat{S}_N(t) \left[1-
\frac{d_\ell}{v} \displaystyle\sum_{n=1}^\infty\,
\displaystyle\sum_{m=0}^n \, s_m^{(n)}  \,\frac{\left(\ln \ln N\right)^m} {\left(\ln N\right)^{n}} \right]
\label{SNtex}
\end{equation}
with
\begin{equation}
\label{hatSNt}
\widehat{S}_N(t)=V_0 (2D)^{d_\ell/2} t^{d_\ell/d_w^{\ell}} \left(\frac{\ln N}{\hat{c}} \right)^{d_{\ell}/v} \;
\end{equation}
and
\begin{eqnarray}
\label{s10}
s_0^{(1)}&=&-\omega  \\
s_1^{(1)}&=&\hat{\mu}   \\
s_0^{(2)}&=&
-(\beta-1) \left( \frac{\pi^2}{12}+\frac{\omega^2}{2} \right) -
(\hat{c} \hat{h_1}-\hat{\mu} \omega) \\
s_1^{(2)}&=& -\hat{\mu}^2 + (\beta-1) \hat{\mu} \omega \\
s_2^{(2)}&=& -\frac{1}{2} (\beta-1) \hat{\mu}^2 \; .
\label{stilde}
\end{eqnarray}
Here $\omega=\gamma+\ln \hat{A} \hat{c}^{\hat{\mu}}$, $\gamma
\simeq 0.577215$ is the Euler constant, and $\beta=d_\ell/
v=d_\ell (d_w^\ell-1)/d_w^\ell$.
Inserting (\ref{s10}) and the definition of
$\omega$ in Eq.\ (\ref{SNtex}) we get an expression for $\langle S_N(t) \rangle $
that almost coincides with that obtained before in Eq.\ \eqref{SNtapprox} by a much simpler analysis.
The only difference between the two expressions is that, in the first-order correction, the term $\ln 2$ plays in Eq.\ (\ref{SNtapprox}) the role of the Euler constant $\gamma$ in Eq.\ (\ref{SNtex}).

\begin{figure}
\includegraphics{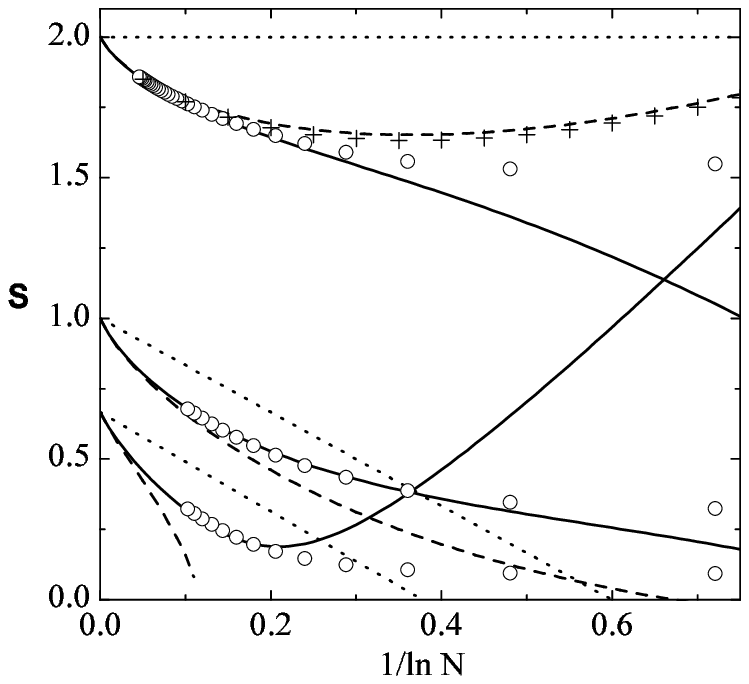}
\caption{\label{figSeucl}
${\bf S}=[S_N(t)/V_0]^{2/d}/(4 D t \ln N)$ versus $1/\ln N$ for, from top to bottom, dimension $1$, $2$ and $3$ and $t=200$ (inside time regime II). We have used $N=2^m$ with $m=3,\cdots,14$ for $d=2,3$, and   $m=3,\cdots,30$ for $d=1$.
The numerical results are plotted
as circles and the broken [solid] lines correspond to the theoretical
predictions for $S_N(t)$  to first [second] order as given by Eq.\ (\protect{\ref{SNt}}).
Notice that the  approximation of order 0 would be a horizontal line (not shown here) passing through $1/d$.
The crosses correspond to the prediction of Sastry and Agmon [Eq.\ (22) of Ref. \protect\cite{Sastry} with $\alpha=1$].
The dotted lines correspond to the result of Larralde {\em et al.} \protect\cite{Larral1} after correcting
the amplitude of the main term (see \protect{\cite{PREeucl}}).
}
\end{figure}

\begin{figure}
\includegraphics{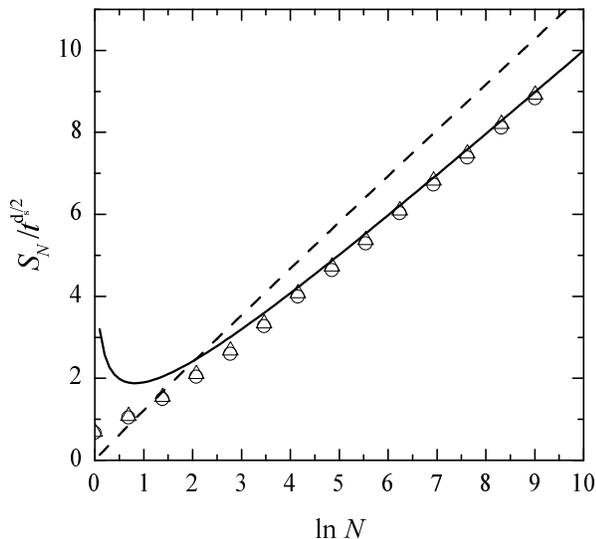}
\caption{\label{figSfrac}
Plot of $S_N(t)/t^{d_{\ell}/d_w^{\ell}}$ versus $\ln N$  in
the two-dimensional incipient percolation aggregate for $N=2^0,2^1,\ldots,2^{13}$. The circles [triangles] are the simulation results for $t=1000$ [$t=500$] averaging over $2000$ aggregate realizations.
The dashed  line is the zeroth-order theoretical prediction with $c=1.05$ and $v=1.7$ and the solid line is the first-order approximation with $c=1.05$, $v=1.7$, $\mu=0.8$ and $A=1$.
}
\end{figure}

In Fig.\ (\ref{figSeucl}) the
first and the second order approximations for $\langle S_N(t) \rangle $ are
compared with Monte Carlo simulation results for $N=2^3$, $2^4$,
\ldots, $2^{14}$ at $t=200$ in the simple cubic Euclidean lattice
with $d=1$, $2$ and $3$.
An excellent agreement was found for $N
\gtrsim 1000$. The uncertainty in the values of the parameters
$\hat{A}$, $\hat{\mu}$, $\hat{c}$ and $\hat{h}_1$ appearing in the
general expression of the point mortality function for fractal
media in Eq.\ (\ref{hpzt}) does not allow a clear comparison in this
case. However, using the tentative values listed in Table
\ref{table2} for the two-dimensional percolation aggregate, good
agreement has been found with Monte Carlo simulation results
\cite{PREfract} as shown in Fig.\ \ref{figSfrac}.

A remarkable fact associated with the asymptotic series
expression (\ref{SNtex}) for $\langle S_N \rangle $ is the large value of the
corrective terms even for very large number of particles $N$. We
have, for example, $\ln \ln N/\ln N \sim 0.075$ for one mol of
random walkers $N \sim 10^{23}$. Hence, these corrective terms
must be included in order to make a sensible estimate of the
territory expected to be covered by any number of random walkers in
simulations or experiments. As discussed in Sec. \ref{sect_2},
there are three time regimes in the territory problem and Eq.
(\ref{SNtex}) is derived only for regime II. Once the trails of
the random walkers no longer overlap (regime III), the series in Eq.
(\ref{SNtex}) fails to converge,  and the convergence condition
$s_m^{(n)}  \left(\ln \ln N\right)^m /\left(\ln N\right)^{n}\ll 1$
determines the span of the regime II. We know that in the
two-dimensional square lattice the parameter $\hat{A}=1/\ln t$
(see Table \ref{table2}) and, consequently, the first corrective
term in the series is comparable with the main term if $\vert \ln
\hat{A} \vert \sim \ln N$. This condition implies a crossover
time $\tau_{\times} \sim e^N$ from regime II to regime III.
Similarly, the crossover time $\tau_{\times} \sim N^2$ is found
for the three-dimensional simple cubic lattice. For the one-dimensional lattice
and any fractal lattice with spectral dimension $d_s < 1$ the
parameter $\hat{A}$ is time independent and regime III is never
reached as we have already discussed in Sect. \ref{sect_2}.
Thus a unified and purely analytical criteria for the time scale
corresponding to the transition between the two regimes II and
III emerges naturally from the series in Eq.\ (\ref{SNtex}).

\subsection{Geometry of the territory explored}
\label{geometry}
\begin{figure*}[t]
\includegraphics[width=17cm]{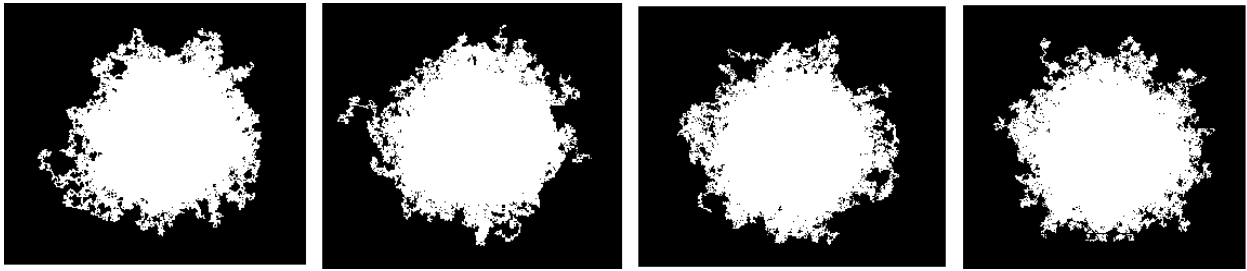}
\caption{\label{SNtsim}
Four successive scaled snapshots of the set of sites visited  by $N=700$ random walkers on the two-dimensional lattice for times (from left to right) $t=2000$, $t=4000$, $t=6000$ and $t=8000$. The second snapshot has been shrunk by the factor 1/$\protect{\sqrt{2}}$, the third  by the factor $1/\protect{\sqrt{3}}$ and the last by the factor $1/2$.
}
\end{figure*}

The asymptotic series in Eq.\ (\ref{SNtex}) can be used
to unveil the geometric properties of the set of visited sites
 on Euclidean lattices \cite{PREeucl}. In Fig. \ref{Territory} one
discerns a compact circular core and a dendritic ring composed of
the trails of those random walkers that have travelled further than the
rest. We define $R_0(N,t)$ as the average distance between the origin and
the nearest unvisited site at time $t$ (the radius of the inner
compact core) and $R_+(N,t)$ as the average maximum distance reached by
any of the $N$ random walkers by time $t$. Assuming compact
exploration in the sense of de Gennes \cite{Gennes},  a reasonable
estimate of $R_0(N,t)$ is given by $\left[ \langle S_N(t) \rangle /V_0
\right]^{1/d}$. Qualitative arguments also yield an approximation
for $R_+(N,t)$ on the $d$-dimensional Euclidean lattice
\cite{PREeucl}. In one dimension, the territory covered by $N$
independent random walkers is obviously a segment stretching from
the maximum span on the left side of the origin to the maximum
span on the right side and, consequently,  we have
$R_+(N,t)=\langle S_N(t) \rangle|_{d=1} /2 \approx \left[ 4 D t \ln N \right]^{1/2}$. The
$d$-dimensional walk of a random walker  over the simple square or cubic
lattice is now decomposed into $d$ orthogonal one-dimensional
random walks and we get $R_+(N,t)\approx \left[4 D (t/d) \ln N
\right]^{1/2}$ because, on average, the random walker  travels along each
direction only the $d$-th part of the time. Taking into account
that on Euclidean lattices $d_{\ell}=d$, $d_w^{\ell}=v=2$ and
$c=d/2$ as shown in Table \ref{table2} we can write the main term
of Eq.\ (\ref{SNtex}) as $\widehat{S}_N  \approx V_0
\left[R_+(N,t)\right]^d$.
Therefore, the thickness of the dendritic layer $R_+(N,t)-R_0(N,t)\approx \left[\langle S_N \rangle/V_0\right]^{1/d}-\left[\widehat{S}_N/V_0\right]^{1/d}$ is given by the corrective terms in Eq.\ (\ref{SNtex}) as
follows: $R_+(N,t)-R_0(N,t) \approx -(\ln \hat{A}(t)/\ln N) R_+(N,t)$.
This means that the dendritic layer's thickness is a
fraction of the compact core radius that grows with time as $\ln
\ln t$ for $d=2$ and $\ln \sqrt{t}$ for $d=3$ (see Table
\ref{table2} for $\hat{A}$). In two dimensions this ratio grows so
slowly that the set of distinct sites visited scales as $\sqrt{t}$, and is almost statistically self-similar in
regime II as Fig.\ \ref{SNtsim} shows.
For the crossover times $\tau_{\times} =\mathcal{O}(e^N)$ ($d=2$) and $\tau_{\times} =\mathcal{O}( N^2)$
($d=3$), the dendritic ring outruns the inner compact core and we
enter into regime III.

\section{Exit times}
\label{exit}
\begin{figure}
\includegraphics{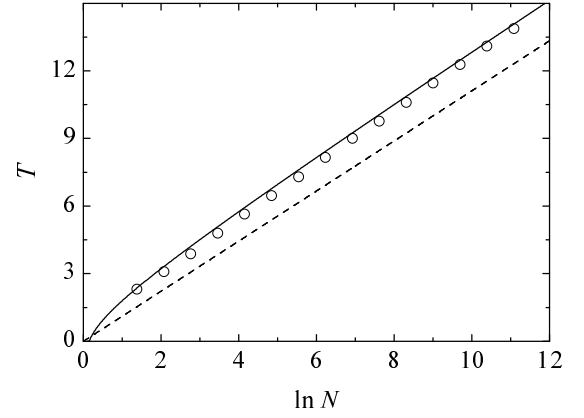}
\caption{\label{t1Nagreg}
Plot of $T\equiv \left[ \langle t_{1,N} \rangle/\ell^{d_w^\ell} \right]^{-\delta}$
 versus $\ln N$ ($N=2^2$, $2^3$, \ldots, $2^{16}$) for the two-dimensional incipient percolation
aggregate where $\delta=1/(d_w^\ell-1)$. The circles are  simulation
results for $\ell=50$ and the lines are  the zeroth-order (broken line) and first-order (solid line) asymptotic approximations with $\hat{c}=0.9$, $\hat{v}=1.714$ and $\hat{\mu}=-0.4$.}
\end{figure}

Now we turn to the order statistics problem of evaluating  the quantities $\langle t_{j,N} \rangle$, $j=1,2,\ldots$,
i.e., the average time (exit time) taken by the $j$-th random walker of a set of $N$ to arrive at a given
border containing the origin site, and its moments. Specific techniques have been
developed for the calculation of these quantities \cite{PRLYus,YusLin,YusAcLin,fptagreg,WSL}.
They will be discussed below, but first it is interesting to explore the connection between
the first passage time and the territory problems.
We have found above that for Euclidean media and also for media with spectral dimension $d_s=2 d_f/d_w < 2$, the random walkers perform an almost compact exploration of the lattice in the sense of de Gennes \cite{Gennes}. This
means the following: If the maximum distance $\ell$ from the origin site is
reached by any of the $N$ random walkers at time $t$, all the sites at a distance
smaller than $\ell$ have already been visited. The territory explored by the
$N$ random walkers is roughly a hypersphere of radius $\ell$, and we can write:
\begin{equation}
\label{SNt1N}
\langle S_N\left[ t_{1,N}(\ell) \right] \rangle \approx V_0 \ell^{d_\ell} \; .
\end{equation}
Taking into account Eqs.\ (\ref{SNtex}), (\ref{hatSNt}) and (\ref{s10}) one easily
finds that Eq.\ (\ref{SNt1N}) implies
\begin{eqnarray}
\label{t1Napp}
t_{1,N} &\approx& \left( \frac{\ell}{\sqrt{2 D}} \right)^{\alpha+1}
            \left( \frac{c}{\ln N} \right)^\alpha  \hfill \, \nonumber \\
&&\times \left\{1+\alpha \frac{\mu \ln \ln N-\gamma-\ln \lambda_0}{\ln N}+\ldots\right\}
\end{eqnarray}
where $\alpha=d_w^\ell-1$ and $\lambda_0=A c^\mu$. It is remarkable that this equation, derived
following simple arguments, is exact to first order, i.e., all terms in Eq.\ \eqref{t1Napp} are exact.
A rigorous approach leads to the result \cite{PRLYus,YusAcLin,fptagreg}:
\begin{eqnarray}
\label{t1Nm}
\langle t_{1,N}^m \rangle &=& \left[ \frac{\ell}{\sqrt{2 D}}
\right]^{m (\alpha+1)}\left[\frac{c}{\ln (\lambda_0 N)}\right]^{m \alpha}
\biggl\{1+\biggr. \nonumber \\
&& + \frac{m \alpha \left(\mu \ln \ln \lambda_0 N-\gamma\right)}{\ln \lambda_0 N}
         +\frac{m \alpha}{2 \ln^2 (\lambda_0 N)} \nonumber \\
&& \times \biggl[(1+ m \alpha)\left( \frac{\pi^2}{6} +\gamma^2 \right)+2 \mu \gamma-2 h_1 c \biggr. \nonumber \\
&& -2 \mu \left( \mu+(1+ m \alpha) \gamma \right) \ln \ln \lambda_0 N \nonumber \\
&& \biggl.+(1+ m \alpha) \mu^2 \ln^2 \ln \lambda_0 N \biggr] \nonumber \\
&& \biggl. + {\cal O} \left(\frac{\ln^3 \ln \lambda_0 N}{\ln^3 \lambda_0 N} \right) \biggr\} .
\end{eqnarray}
\begin{figure}[t]
\begin{center}
\leavevmode
\epsfxsize = 5.7cm
\epsffile{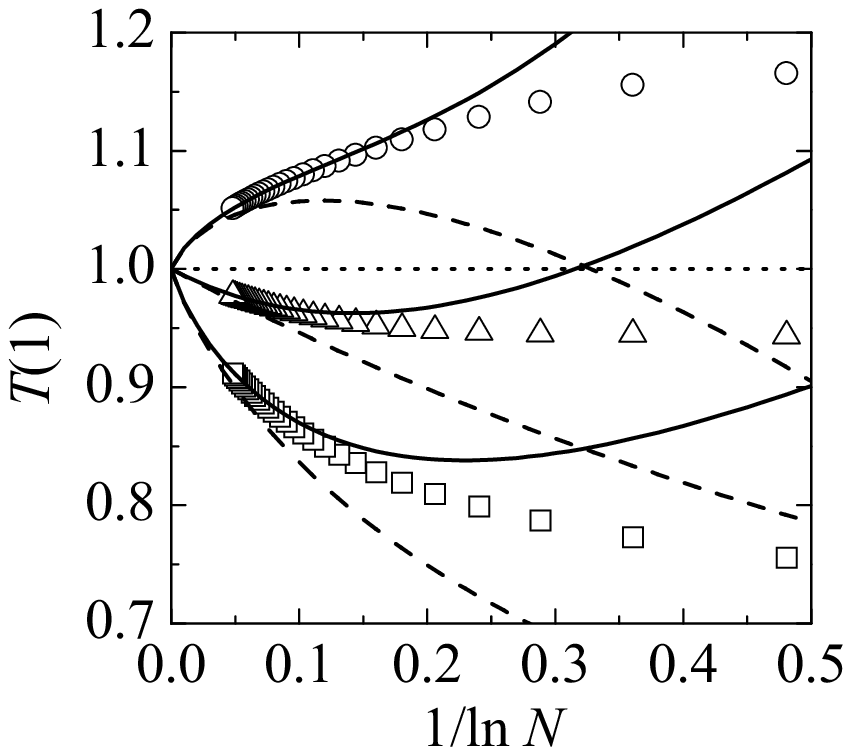}
\leavevmode
\epsfxsize = 5.7cm
\epsffile{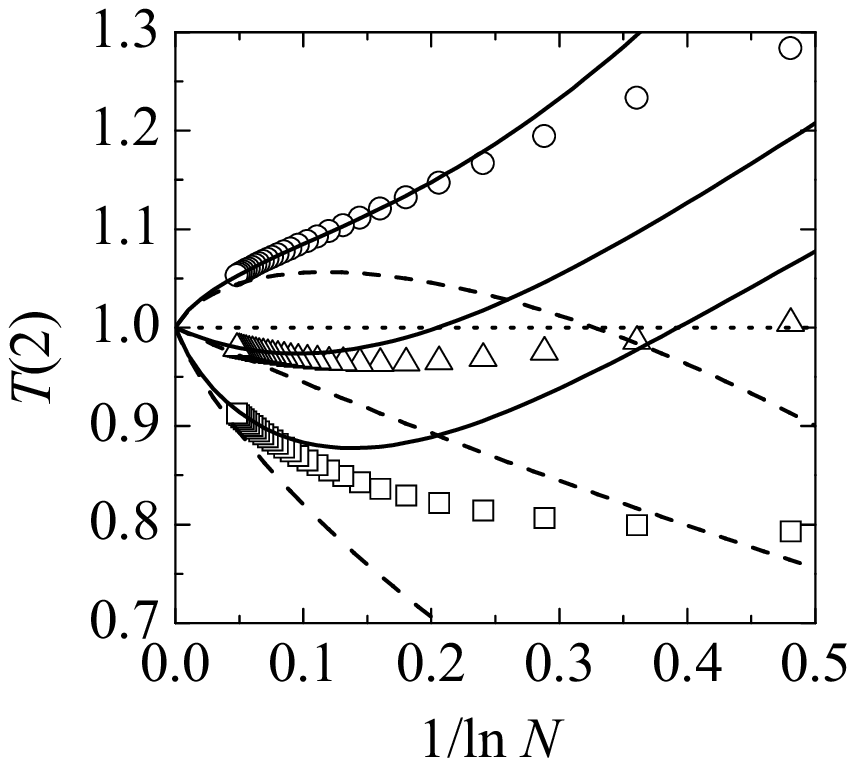}
\leavevmode
\epsfxsize = 5.7cm
\epsffile{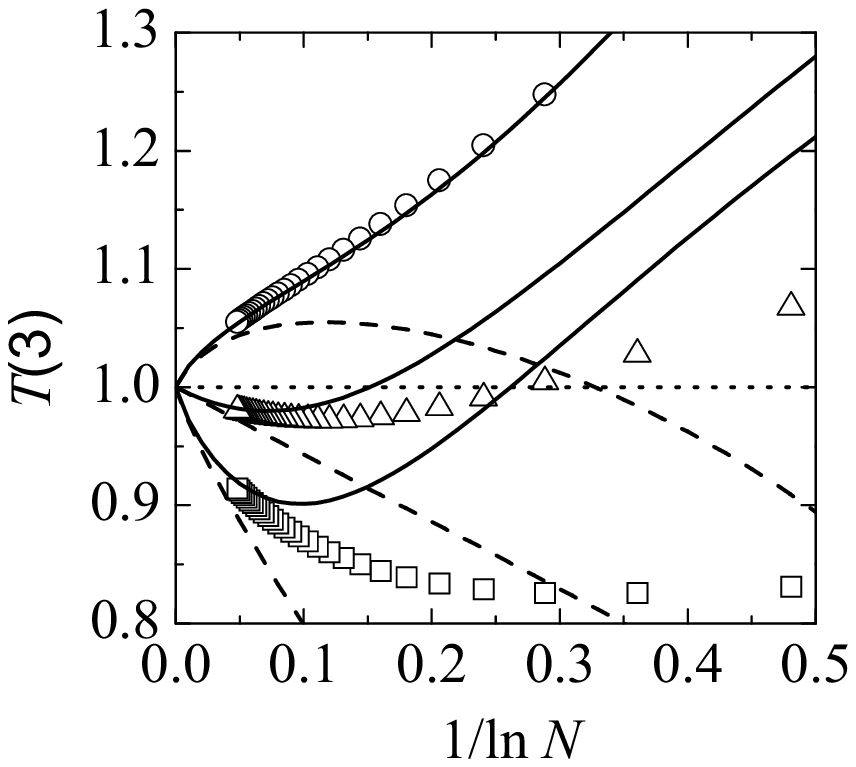}
\end{center}
\caption
{Scaled moments $T(m)=(2d/r^2)\ln (\lambda_0N)\langle t^m_{1,N}\rangle^{1/m}$ for the first arrival of the first of $N$ particles at a prescribed boundary as a function of $1/\ln N$.  First panel: first
moment, $m=1$ (i. e., mean first passage time); second panel: second moment, $m=2$;
third panel: third moment, $m=3$.  Numerical results are indicated by circles
($d=1$), triangles ($d=2$), and squares ($d=3$).  Asymptotic results (cf.
Eq.~(\ref{t1Nm})) to zeroth order, first order, and second order are
shown by dotted curves, dashed curves, and solid curves respectively.
}
\label{t1Neucld}
\end{figure}
If we write the series in Eq.\ (\ref{t1Nm}) in terms of $\ln N$ instead of
$\ln \lambda_0 N$ we find that it coincides, to first order, with
the approximation in Eq.\ (\ref{t1Napp}). The moments of $\langle t_{j,N}^m \rangle$
for $j > 1$ have also been calculated \cite{PRLYus,YusAcLin,fptagreg} and the
result is
\begin{eqnarray}
\label{tjNm}
\langle t_{j,N}^m \rangle &=&
\langle t_{1,N}^m \rangle \hfill \nonumber \\
&&+\frac{m \alpha c^{m \alpha} \ell^{m(\alpha+1)}}{\left[\sqrt{2 D}\ln (\lambda_0 N)\right]^{m \alpha+1}} \sum_{n=1}^{j-1}
\frac{\Delta_n}{n}
\end{eqnarray}
where $j=2,3,\ldots$ and
\begin{eqnarray}
\label{Dnalpha}
\Delta_n&=&1+\frac{m \alpha+1}{\ln \lambda_0 N} \left[
(-1)^n \frac{S_n(2)}{(n-1) !}+\mu \ln \ln \lambda_0 N  \right.\nonumber \\
&&\left. -\frac{\mu}{m \alpha+1}-\gamma \right]+{\cal O}\left( \frac{\ln^2 \ln \lambda_0 N}{\ln^2 \lambda_0 N}
\right)\; .
\end{eqnarray}
The quantities $S_n(i)$ are the Stirling numbers of the first
kind \cite{AbSt}. The expressions (\ref{t1Nm}) and (\ref{tjNm})
are also valid for Euclidean boundaries with minor and obvious changes ($d_w^\ell\rightarrow d_w$, $\ell \rightarrow r$, \ldots).
In Euclidean lattices and deterministic fractals there is no
difference between using chemical or Euclidean distances.
In Fig.\ \ref{t1Neucld},  exact scaled results for the $N$
dependence of $\langle t_{1,N}^m \rangle^{1/m}$ are plotted
against the predictions of Eq.\ (\ref{t1Nm}) for the
one-dimensional lattice, the two-dimensional square lattice and
the three-dimensional simple cubic lattice. Agreement is also
very good for disordered media such as the two-dimensional
incipient percolation aggregate, as shown in Fig.\ (\ref{t1Nagreg}).

\section{Trapping times}
\label{trapping}
\subsection{Lifetime of the first trapped particle}
\begin{figure}
\includegraphics{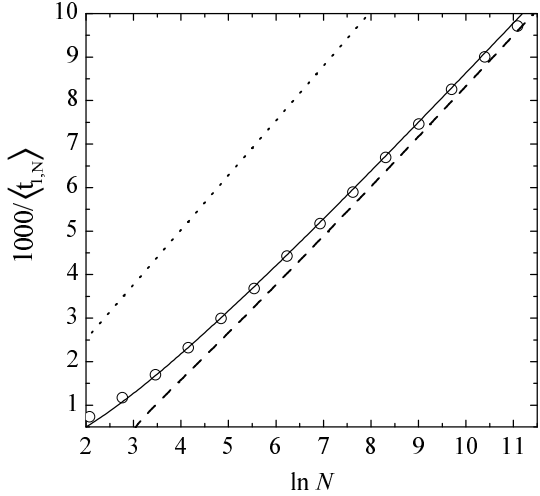}
\caption{\label{t1Ncfig}
The function $10^3/\langle  \mathbf{t}_{1,N}\rangle$ versus $\ln N$ for the
two-dimensional lattice  with  $c=4\times 10^{-4}$ and
$N=2^3,2^4,\ldots,2^{16}$. The simulation results are averaged over
$10^5$ configurations (circles).  The dotted line represents the asymptotic
approximation of order 0.
We also plot the results obtained by numerically integrating Eq.\ (\ref{t1Nmtrap}) when the first-order (dashed line) and second-order (solid
line) asymptotic approximations for $\langle S_N(t) \rangle$ are used.
}
\end{figure}

The ``trapping'' problem is a fundamental problem of random walk theory with
a long tradition \cite{HughesWeiss,SHDBA}. In this problem a lattice is randomly
filled with static traps placed at certain sites, and a particle performs a random
walk on the lattice until it arrives at a site occupied by a trap, when it is absorbed.
Most works have analyzed the survival probability of the random walker, $\Phi_1(t)$, defined
as the probability of the random walker  not being absorbed in the time interval $[0,t]$ \cite{Zumofen}. This problem
have also been generalized to $N > 1$ random walkers, and the survival probability of the set of $N$ random walkers
has been defined  analogously \cite{HalfLine}. In this spirit,  we define
the survival probability $\Phi_N(t)$ that no particle of the initial set of $N$ diffusing particles has
been trapped by time $t$. This survival function is given by
\begin{equation}
\label{phiNt}
\Phi_N(t)=\langle (1-c)^{S_N(t)} \rangle \; ,
\end{equation}
where $c$ is the concentration of traps and the angle brackets denote an average over all
realizations of the $N$ particle random walk on the lattice. Usually, $\Phi_N(t)$ has been estimated
using the Rosenstock approximation \cite{HughesWeiss,SHDBA,HalfLine,TrapOrder,Hollander} in
which the average of the exponential is identified with the exponential of the average:
\begin{equation}
\label{Rosens0}
\Phi_N(t)=e^{-\lambda \langle S_N(t) \rangle}\; ,
\end{equation}
with $\lambda=-\ln (1-c)$. Higher order terms are calculated in the extended
Rosenstock approximation (or truncated cumulant expansion) first proposed by
Zumofen and Blumen \cite{Zumofen}. The first-order Rosenstock approximation
is \cite{TrapOrder}
\begin{equation}
\label{Ros:first}
\Phi_N^{(1)}=\exp\left[-\lambda \left\langle S_N \right\rangle
\left(1+\displaystyle\frac{\lambda}{2}
\frac{\text{Var}(S_N)}{\langle S_N \rangle} \right) \right].
\end{equation}
This means that the error made by using the zeroth-order Rosenstock
approximation is $\mathcal{O}(\lambda^2 \text{Var}(S_N))$. Consequently, this
approximation performs well if the condition $\lambda^2 \text{Var}(S_N) \ll 1$
is satisfied. In Ref. \cite{TrapOrder}, it  was shown that $\text{Var}(S_N) =\mathcal{O}\left(
t^d (\ln N)^{d-2}\right)$ on the simple Euclidean $d$-dimensional lattice, and we have
that the zeroth-order Rosenstock approximation works well when
$\lambda^2 t^d (\ln N)^{d-2} \ll 1$. Hence, the approximation (\ref{Rosens0})
becomes poorer as $N$ increases for $d=3$, and  also for long times when, eventually,
the Donsker-Varadhan regime settles in \cite{DV}, $\Phi_N(t) \sim \exp(-t^{d/(d+2)})$. If we
now define $h_{1,N}(t)=-d \Phi_N(t)/d t$ as the probability that the first absorbed particle of
the initial set of $N$ disappears during the time interval $(t,t+d t]$, it is clear
that, using Eq.\ (\ref{Rosens0}), the average time (lifetime) at which the first particle is trapped is given by
\begin{eqnarray}
\langle \mathbf{t}_{1,N}^m \rangle &=& \displaystyle\int_0^\infty\, t^m h_{1,N}(t)\, d t \nonumber \\
&\simeq& m \displaystyle\int_0^\infty\, t^{m-1} \exp[-\lambda\langle S_N \rangle]\, dt\; .
\label{t1Nmtrap}
\end{eqnarray}
Inserting the main asymptotic term of $\langle S_N\rangle$ given by Eq.\ \eqref{hatSNt} into
Eq.\ (\ref{SNtex}) for the number of distinct sites visited into Eq.\ (\ref{t1Nmtrap})
a zeroth-order approximation for the $m$th moment of the first trapping time, $\mathbf{t}_{1,N}$ was
found in Ref. \cite{TrapOrder} as follows:
\begin{equation}
\langle \mathbf{t}_{1,N}^m \rangle \simeq
\frac{\Gamma(1+2m/d)}{\left(\lambda V_0 \right)^{2m/d}} \frac{d^m}{(4D\ln N)^m}\; .
\label{t1Nc}
\end{equation}
No further terms in the expansion of $\langle S_N \rangle $ were considered because, in the two-
and three-dimensional Euclidean lattice case, they depend on time,  and the analytical
integration in Eq.\ (\ref{t1Nmtrap}) cannot be explicitly performed.
In Fig.\ \ref{t1Ncfig},  numerical simulation results for $\langle \mathbf{t}_{1,N} \rangle$ on the
two-dimensional square lattice with $c=4\times 10^{-4}$ are compared with the predictions of
Eq.\ (\ref{t1Nc}) and those of Eq.\ (\ref{t1Nmtrap}) obtained by numerical integration
using the first and second order corrective terms of $\langle S_N \rangle $ in Eq.\ (\ref{SNtex}).

\subsection{Lifetime of the $j$th trapped particle}
\begin{figure}
\includegraphics{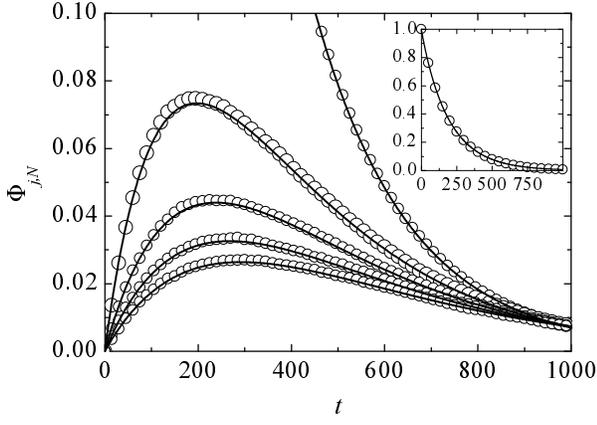}
\caption{\label{PhijNd2}
The $j$th survival probability $\Phi_{j,N}$ versus time $t$ for (from top to bottom)
$j=0,1,2,3,4$, with $N=1000$ and $c=4\times10^{-4}$ for the two-dimensional lattice.
The lines represent $\Phi_{j,N}^{(02)}(t)$, i.e., the zeroth-order Rosenstock
approximation with $\langle S_N(t)\rangle$ given by the second-order
asymptotic approximation. The circles are simulation results averaged over $10^6$
configurations. Inset: $\Phi_{0,N}(t)$.}
\end{figure}

Until now we have discussed the statistics of the absorption of the first particle, but
the problem of the order statistics have also been solved for $ 1 < j \ll N$ \cite{TrapOrder}. This
has been possible for independent random walkers. If we define $\Psi(t)$ as the
survival probability of a single random walker in a given trapping configuration, then
the distribution for the absorption of $j$ particles is given by
\begin{eqnarray}
\Psi_{j,N}(t)&=& \binom{N}{j}
 \left(1-\Psi\right)^j  \Psi^{N-j} \nonumber\\
 &=&
\binom{N}{j}
\sum_{m=0}^j (-1)^m
\binom{j}{m}
\Psi^{N-j+m}.
\label{PsijN}
\end{eqnarray}
An average over all different trap configurations yields $\Phi_N(t)=\langle
\Psi^N(t) \rangle$, the survival probability of the $N$ random walkers and $\Phi_{j,N}(t)=
\langle \Psi_{j,N}(t)\rangle$, the probability that exactly $j$ random walkers from
the set of $N$ have been trapped in the time interval $[0,t]$. From Eq.\ (\ref{PsijN}) we
get
\begin{eqnarray}
\label{PhijNder}
\Phi_{j,N}(t)
&=&(-1)^j \binom{N}{j} \Delta^j \Phi_N (t) \\
&\simeq& (-1)^j \binom{N}{j} \displaystyle\frac{d^j}{d N^j} \Phi_N(t)  \; ,
\end{eqnarray}
using the backward difference formula for the $j$th derivative and the approximation
$\Delta^j \simeq d^j/d N^j$ when $j \ll N$. In Fig.\ (\ref{PhijNd2}) the $j$th
survival probability $\Phi_{j,N}(t)$ obtained from Eq.\ (\ref{PhijNder}) using
the Rosenstock approximation for $\Phi_N(t)$ is compared with simulation results
for the two-dimensional lattice with a concentration of traps $c=4\times 10^{-4}$.

\begin{figure}
\includegraphics{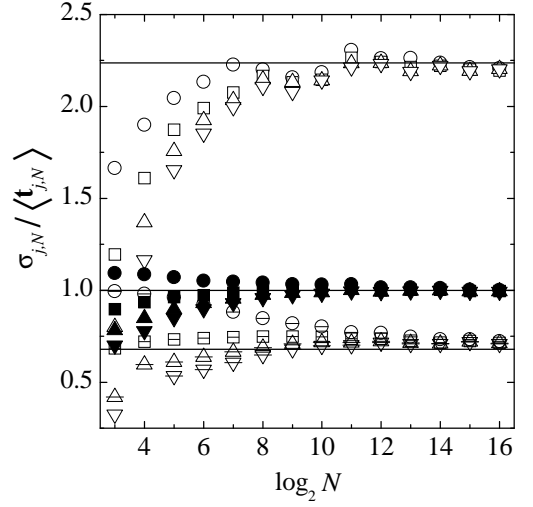}
\caption{\label{sratio}
The ratio $\sigma_{j,N}/\langle  \mathbf{t}_{j,N}\rangle$ , $j=1$ (circles)
$j=2$ (squares), $j=3$ (up triangles)  $j=4$ (down triangles),  $N=2^3,
2^4,\ldots,2^{16}$, for $d=1$ with $c=8\times 10^{-3}$ (hollow
symbols at the top of the figure),  $d=2$ with $c=4\times
10^{-4}$ (filled symbols) and $d=3$ with  $c=4\times 10^{-5}$
(symbols with a bar at the bottom of the figure).
The simulation results are averaged over $10^5$ configurations for $d=1$ and $d=2$,
and over  $10^4$ configurations for $d=3$.
The  lines represent the (main order) asymptotic theoretical results, namely,
$\sqrt{5}$ for $d=1$,   $1$ for $d=2$ and $0.678968\cdots$ for $d=3$.}
\end{figure}

The probability that the $j$th absorbed particle of the initial set of $N$ disappears
during the time interval $(t,t+d t]$, $h_{j,N}(t)$, satisfies the recurrence
relation:
\begin{equation}
\label{hjN}
h_{j+1,N}(t)=h_{j,N}(t)-\frac{d}{d t} \Phi_{j,N}(t)\; ,
\end{equation}
with $h_{0,N}(t)$. The $m$th moment of the time at which the $j$th particle is
trapped is given by
\begin{equation}
\label{tjNmtrap}
\langle \mathbf{t}_{j,N}^m \rangle=\displaystyle\int_0^\infty\, t^m h_{j,N}(t) d t\; ,
\end{equation}
and, taking into account Eqs. (\ref{hjN}) and (\ref{PhijNder}), we finally
arrive at a recurrence relation for these moments
\begin{equation}
\langle \mathbf{t}_{j+1,N}^m \rangle =
\langle \mathbf{t}_{j,N}^m \rangle +
(-1)^j
\left(
\begin{array}{c}
 N \\ j
\end{array}
\right)
 \Delta_j \langle \mathbf{t}_{1,N}^m \rangle  \;  .
\label{tjNb}
\end{equation}
For large $N$ and small $j$, this relation yields
\begin{equation}
\langle \mathbf{t}_{j+1,N}^m \rangle \simeq
\langle \mathbf{t}_{j,N}^m \rangle+
m \frac{d^m \Gamma(1+2m/d)}{\left(\lambda V_0 \right)^{2m/d}(4D)^m}
\frac{(\ln N)^{-1-m}}{j}\; ,
\label{tjNc}
\end{equation}
where the difference operator $\Delta^j$ has been approximated by $d^j/d N^j$
and $\langle \mathbf{t}_{1,N}^m \rangle$ was given by Eq. (\ref{t1Nc}). It is remarkable
that the main asymptotic term of the ratio $\sigma_{j,N}/\langle \mathbf{t}_{j,N} \rangle$
between the variance $\sigma_{j,N}=\sqrt{\langle \mathbf{t}_{j,N}^2 \rangle-\langle
\mathbf{t}_{j,N} \rangle^2}$ and $\langle \mathbf{t}_{j,N} \rangle$ depends only on the dimension
of the lattice as is easily shown from Eq. (\ref{tjNc}):
\begin{equation}
\frac{\sigma_{j,N}}{\langle \mathbf{t}_{j,N} \rangle}  \simeq
\frac{\left[\Gamma(1+4/d)- \Gamma^2(1+2/d)\right]^{1/2}}{\Gamma(1+2/d)}
\label{ratiostjN}
\end{equation}
In Fig.\ \ref{sratio} this ratio is plotted for $d=1$, $2$ and $3$
for several values of $j$ and $N$ and compared with simulation
results. Good agreement is found for $N \gtrsim 1000$.

\subsection{One-dimensional case: rigorous results}
For the one-dimensional lattice one can obtain rigorous order-statistics asymptotic expressions for the trapping times (lifetimes) from the knowledge of the order statistics of the diffusion process in the presence
of two fixed traps \cite{TrapOrder}, without resorting to the Rosenstock approximation.
We only quote here the results up to first-order corrective terms for large $N$ (second-order terms were calculated in \cite{TrapOrder}):
\begin{equation}
\langle \mathbf{t}_{j,N}^m \rangle   =
 \frac{\Gamma(1+2m)}{(2\lambda)^{2m}}
 \frac{\tau_{j,N}(m)}{\left( 4D\ln \kappa N  \right)^m}
\label{tmjN1Dasin}
\end{equation}
with $ \tau_{j,N}(m)= \tau_{1,N}(m)+ \delta_{j,N}(m)$,
\begin{equation}
\tau_{1,N}(m) = 1+m \frac{\frac{1}{2}\ln\ln\kappa  N-\gamma}{\ln \kappa  N} +
         {\cal O}\left(\frac{\ln^2\ln\kappa  N}{\ln^2\kappa  N}\right) ,
\label{tau1N}
\end{equation}
\begin{equation}
\delta_{j,N}(m)= \frac{m}{\ln \kappa N}\;
 \sum_{n=1}^{j-1} \frac{1}{n}  ,
\label{deltamjN}
\end{equation}
and $\kappa=1/\sqrt{\pi}$.

\section{SUMMARY AND OPEN PROBLEMS}
\label{sect_5}
In this review we have discussed some recent advances in the field of multiparticle
independent random walks.
The random walk is one of the simplest non-equilibrium models in statistical
physics, and has played an important role in the study of transport troughout the
past century \cite{Einstein,HughesWeiss,SHDBA}. Problems concerning a single
random walker have been the subject of thorough study, and it is today a textbook discipline \cite{HughesWeiss}.
However, multiparticle random walk problems have
been posed only very recently \cite{Interacting,Larral1,Havlin,fewN} and we are just starting to understand
what happens in this case. We have here dealt exclusively with independent random
walkers all starting from the same site on Euclidean, deterministic fractal and
disordered lattices. Three problems have been discussed:

\smallskip
(a) The territory problem: the estimate of the average number  $\langle S_N(t) \rangle $ of distinct sites  visited at time $t$ by $N$ random walkers starting from the origin at $t=0$.

(b) Order statistics of exit times  from a (hyper)spherical region: the
evaluation of the moments of the  time $t_{j,N}$ taken by the $j$-th, $j=1$, $2$, \ldots, random walker
out of $N$ to arrive at a closed boundary at a given distance from the origin.

(c) Order statistics of trapping times: the estimate of the moments of the elapsed time $\mathbf{t}_{j,N}$ until the first $j$ particles from a set of $N$ are trapped in a disordered configuration of trapping sites.
\smallskip

These problems can not be solved by simple generalizations of the solution of the $N=1$ case, and
specific techniques hava had to be  developed for the multiparticle problems even for independent random walkers.
It is remarkable that these techniques lead to solutions of the aforementioned problems in terms of asymptotic series for large $N$ which share the same mathematical form \cite{PRLYus,PREeucl,PREfract,YusAcLin,HalfLine,TrapOrder,fptagreg}.
The corrective terms of these series decay logarithmically in the number $N$ of random walkers, and are consequently, very important even when $N$ is very large. This is reflected in the large magnitude of the fluctuations on the diffusion front (see Sec.\ \ref{geometry}).

Despite the success in the calculation of the territory explored (c.f. Sec.\  \ref{territory}),
the order statistics of the exit times (c.f. Sec.\ \ref{exit}) and the order statistics of the trapping time (c.f.  Sec.\ \ref{trapping}),  there are still many open problems.
For example, there exists the problem of the evaluation of the distribution of the
territory explored or, equivalently, the moments $\langle S_N^m (t) \rangle$, $m=2$, $3, \cdots$
Only the moments of the number of distinct sites visited in a {\em given} direction of
a one-dimensional lattice up to time $t$ have been calculated rigorously \cite{HalfLine}.
There are good reasons to believe that these moments may be expressed by series in $\ln N$ of the
same form as those in Eq.\ (\ref{SNtex}), but this is only a conjecture for general
media \cite{TrapOrder}.
The development of a procedure for the calculation of these
moments in Euclidean and fractal lattices is a still a challenge.
The work discussed in this review could also be extended in other directions.
For example, the extension of our work to the case in which the traps are randomly distributed on a fractal (deterministic or stochastic) is an interesting open problem.
Another front opened recently is the study of unknotting in granular chains on
a vibrating plate \cite{Gchains}. The time corresponding to the unknotting of the
first knot in a chain with $N$ knots corresponds to the first passage time of a
random walker of a set of $N$ random walkers interacting through an excluded volume pair
potential. This is a good motivation, among many others, to investigate the influence
of interactions in the quantities we have considered in this review, namely, $\langle S_N(t) \rangle $, $\langle t_{j,N}^m \rangle$ and $\langle \mathbf{t}_{j,N}^m \rangle$.

\acknowledgments
This work has been supported by the Ministerio de Ciencia y Tecnolog\'{\i}a (Spain) through Grant No. BFM2001-0718.




\begin{thebibliography}{99}


\bibitem{Einstein} J. Stachel and D. C. Cassidy, Eds., {\em The Collected
Papers of Albert Einstein}, Vol. 2, (Princeton University Press, Princeton, 1990).

\bibitem{HughesWeiss} B. H. Hughes, Ed., {\em Random Walks and Random Environments, Volume 1: Random
Walks} (Clarendon Press, Oxford, 1995); {\em Random Walks and Random Environments, Volume 2:
Random Environments} (Clarendon Press, Oxford, 1996).
G. H. Weiss, {\em Aspects and Applications of the Random
Walk} (North-Holland, Amsterdam, 1994).
\bibitem{SHDBA} S. Havlin and D. Ben-Avraham, Adv. Phys. {\bf 36}, 695 (1987), and references
therein.
\bibitem{Scheidegger} A. E. Scheidegger, {\em The Physics of Flow through Porous Media}, (University
of Toronto Press, Toronto, 1974).

\bibitem{BundeHav} A. Bunde and S. Havlin, Eds., {\em Fractals in Science}, (Springer-Verlag, Berlin, 1994).

\bibitem{Interacting} M. E. Fisher, J. Stat. Phys. {\bf 34}, 669 (1984). C. Aslangul, J. Phys. A {\bf 32}, 3993 (1999). V. Kukla, J. Kornatowsky, D. Demuth, I. Girnus, H. Pfeifer, L. V. C. Rees, S. Schunk, K. K. Unger and J. Karger, Science {\bf 272}, 702 (1996).

\bibitem{Larral1} H. Larralde, P. Trunfio, S. Havlin, H. E. Stanley, and
G. H. Weiss, Nature (London) {\bf 355}, 423 (1992); Phys. Rev. A {\bf 45}, 7128 (1992).

\bibitem{Havlin} S. Havlin, H. Larralde, P. Trunfio, J. E. Kiefer, H. E. Stanley and G. H. Weiss, Phys. Rev. A. {\bf 46}, R1717 (1992).

\bibitem{Shlesinger}  M. F. Shlesinger, Nature (London) {\bf 355}, 396 (1992).

\bibitem{fewN}  G. H. Weiss, K. E. Shuler, and K. Lindenberg, J. Stat. Phys.
{\bf 31}, 255 (1983).

\bibitem{BD} A. Bunde and J. Dr\"ager, Physica A {\bf 202}, 371 (1994).
\bibitem{PRLYus} S. B. Yuste, Phys. Rev. Lett. {\bf 79}, 3565 (1997); Phys. Rev. E {\bf 57}, 6327 (1998).
\bibitem{YusLin} S. B. Yuste and K. Lindenberg, J. Stat. Phys. {\bf 85}, 501 (1996); S. B. Yuste, Phys. Rev. E {\bf 57}, 6327 (1998).
\bibitem{DK} J. Dr\"ager and J. Klafter, Phys. Rev. E {\bf 60}, 6503 (1999).
\bibitem{PREeucl} S. B. Yuste and L. Acedo, Phys. Rev. E {\bf 60}, R3459 (1999); {\bf 61}, 2340  (2000).
\bibitem{JPAYusAc} S. B. Yuste and L. Acedo,  J. Phys. A {\bf 33}, 507 (2000).
\bibitem{PREfract} L. Acedo and S. B. Yuste, Phys. Rev. E {\bf 63}, 011105 (2001).
\bibitem{YusAcLin} S. B. Yuste, L. Acedo and K. Lindenberg, Phys. Rev. E {\bf 64}, 052102 (2001).
\bibitem{KR} P. L. Krapivsky and S. Redner, J. Phys. A: Math. Gen. {\bf 29}, 5347 (1996); Am. J. Phys. {\bf 67}, 1277 (1999).
\bibitem{HalfLine} S. B. Yuste and L. Acedo, Physica A {\bf 297}, 321 (2001).
\bibitem{TrapOrder} S. B. Yuste and L. Acedo, Phys. Rev. E {\bf 64}, 061107 (2001).
\bibitem{SingMol} See, for example, the section ``Single Molecules'' in  Science {\bf 283}, 1667-95 (1999).
\bibitem{weitzlab} M. T. Valentine,  P. D. Kaplan, D. Thota, J. C. Crocker, T. Gisler, R. K.
Prud'homme, M. Beck and D. A. Weitz, Phys. Rev. E {\bf 64}, 061506 (2001).
\bibitem{Broeck} C. Van den Broeck, Phys. Rev. Lett. {\bf 62}, 1421 (1989); Phys. Rev. A {\bf 40}, 7334 (1989).
\bibitem{Stauffer} D. Stauffer and A. Aharony, {\em Introduction to Percolation Theory}, 2nd ed. (Springer-Verlag, Berlin, 1996).
\bibitem{Fractals} A. Bunde and S. Havlin, Eds., {\em Fractals and Disordered Systems}, (Springer-Verlag, Berlin, 1996).
\bibitem{Leath} P. L. Leath, Phys. Rev. B {\bf 14}, 5046 (1976).
\bibitem{DE} A. Dvoretzky and P. Erd\"os, in {\em Proceedings of the Second
Berkeley Symposium on Mathematical Statistics and Probability}, (University
of California Press, Berkeley, 1951).
\bibitem{ABN} B. C. Arnold, N. Balakrishnan and H. N. Nagaraja, {\em A First Course
in Order Statistics}, (John Wiley \& Sons, New York, 1992).
\bibitem{Lindenberg} K. Lindenberg, V. Seshadri, K. E. Shuler, and G. H. Weiss,
J. Stat. Phys. {\bf 23}, 11 (1980).
\bibitem{Hollander} F. den Hollander, G. H. Weiss, in: G. H. Weiss (Ed.), Contemporary
Problems in Statistical Physics, SIAM, Philadelphia, 1994.
\bibitem{Beeler} J. R. Beeler, Phys.  Rev. {\bf 134}, 1396 (1964).
\bibitem{Rosens} H. B. Rosenstock, Phys. Rev. {\bf 187}, 1166 (1969).
\bibitem{Damask}  P. Damask and P. Dienes, {\em Point Defects in Metals},
(Gordon and Breach, New York, 1964).
\bibitem{Oshanin}  G. Oshanin, S. Nechaev, A. M. Cazabat and M. Moreau
Phys. Rev. E {\bf 58}, 6134 (1998); S. Nechaev, G. Oshanin and A. Blumen,
J. Stat. Phys. {\bf 98}, 281 (2000).
\bibitem{Miyagawa}  H. Miyagawa, Y. Hiwatari, B. Bernu and J. P. Hansen,
J. Chem. Phys. {\bf 88}, 3879 (1988); T. Odagaki, J. Matsui and Y. Hiwatari,
Phys. Rev. E {\bf 49}, 3150 (1994).
\bibitem{Martinez} H. L. Martinez, J. M. R. Parrondo and K. Lindenberg, Phys. Rev. E {\bf 48}, 3545 (1993); {\bf 48}, 3556 (1993).
\bibitem{Bidaux} R. Bidaux, J. Chave and R. Vocka, J. Phys. A: Math. Gen. {\bf 32}, 5009 (1999).
\bibitem{fptagreg} L. Acedo and S. B. Yuste, in preparation.
\bibitem{YusteJPA} S. B. Yuste, J. Phys. A {\bf 28}, 7027 (1995).
\bibitem{Sastry}G. M. Sastry and N. Agmon, J. Chem. Phys. {\bf 104}, 3022 (1996);
\bibitem{Gennes} P.-G. de Gennes, C. R. Acad. Sci., Ser. 1 {\bf 296}, 881 (1983); J. Chem.
Phys. {\bf 76}, 3316 (1982).
\bibitem{WSL} G. H. Weiss, K. E. Shuler, and K. Lindenberg, J. Stat. Phys. {\bf 31}, 255 (1983).
\bibitem{AbSt} {\em HandBook of Mathematical Functions}, edited by M. Abramowitz
and I. Stegun (Dover, New York, 1972).
\bibitem{Zumofen} G. Zumofen and A. Blumen, Chem. Phys. Lett. {\bf 83}, 372 (1981).
\bibitem{DV} D. V. Donsker and S. R. S. Varadhan, Commun. Pure Appl. Math. {\bf 28} 525
(1975); P. Grassberger and I. Procaccia, J. Chem. Phys. {\bf 77} 6281 (1982).
\bibitem{Gchains} E. Ben-Naim, Z. A. Daya, P. Vorobieff and R. E. Ecke, Phys. Rev. Lett. {\bf 86}, 1414 (2001).
\end{thebibliography}
\end{document}